\documentclass[iop, apj]{emulateapj}
\usepackage{graphicx}

\begin{document}

\def\etal{et al.\ \rm}
\def\ba{\begin{eqnarray}}
\def\ea{\end{eqnarray}}
\def\etal{et al.\ \rm}
\def\Fdw{F_{\rm dw}}
\def\Tex{T_{\rm ex}}
\def\Fdis{F_{\rm dw,dis}}
\def\Fnu{F_\nu}
\def\FJ{F_{J}}
\def\FJE{F_{J,{\rm Edd}}}

\newcommand\tna{\,\tablenotemark{a}}

\title{How to build Tatooine: reducing secular excitation 
in {\it Kepler} circumbinary planet formation}

\author{Roman R. Rafikov\altaffilmark{1}}
\altaffiltext{1}{Department of Astrophysical Sciences, 
Princeton University, Ivy Lane, Princeton, NJ 08540; 
rrr@astro.princeton.edu}


\begin{abstract}
Circumbinary planetary systems recently discovered by {\it Kepler} 
represent an important testbed for planet formation theories. 
Planetesimal growth in disks around binaries has been expected to 
be inhibited interior to $\sim 10$ AU by secular excitation of high 
relative velocities between planetesimals, leading to their 
collisional destruction (rather than agglomeration). Here we 
show that gravity of the gaseous circumbinary disk in which 
planets form drives fast precession of both the planetesimal 
and binary orbits, resulting in strong suppression of 
planetesimal eccentricities beyond 2-3 AU and making possible 
growth of $1-10^2$ km objects in this region. The precise location
of the boundary of accretion-friendly region depends on the size
of the inner disk cavity cleared by the binary torques and on 
the disk mass (even $0.01$ M$_\odot$ disk strongly 
suppresses planetesimal excitation), among other things. 
Precession of the orbit of the central binary, 
enhanced by the mass concentration naturally present at the 
inner edge of a circumbinary disk, plays key role in this 
suppression, which is a feature specific to the circumbinary 
planet formation.    
\end{abstract}



\section{Introduction.}  
\label{sect:intro}


One of the most intriguing findings of the {\it Kepler}
mission is the discovery of circumbinary planets around 
a sample of close binaries, affectionately termed 
{\it Tatooines} (Doyle \etal 2011; Welsh \etal 2012; Orosz 
\etal 2012a,b; Schwamb \etal 2012). Six such systems are 
known at the moment, with one of them, Kepler-47, harboring 
two planets (see Table \ref{table2}). At least some of the 
{\it Kepler} circumbinary planets are likely 
in the Saturn or Jupiter mass regime. Semi-major axes of 
their orbits are typically small, $a_{pl}\lesssim 1$ AU, 
and close to the limit of dynamical stability 
(Holman \& Wiegert 1999).

These systems provide interesting targets for testing our
understanding of planet formation. It is generally agreed 
that in-situ formation of such planets is impossible because
of the strong dynamical excitation due to central binary at 
their present locations (Meschiari 2012a; Paardekooper \etal 
2012). The question typically addressed 
is at what separation could these planets form, 
subsequent to which they have migrated in. The bottleneck 
for planet formation here is the growth of $1-10^2$ km
planetesimals, which is impossible if planetesimals
have high eccentricities --- instead of merging they get 
destroyed in high-speed collisions. Since the 
excitation by the binary is a rather long-range effect 
(in conventional secular theory eccentricity decays with 
distance only as $r^{-1}$, see equation (\ref{eq:eA})), 
it might easily be the case that the conditions for 
planetesimal growth are realized only beyond $10$ AU 
(Moriwaki \& Nakagawa 2004, hereafter MN04; Scholl \etal 2007).
However, at such separations the timescale for growing
massive cores capable of forming giant planets by core 
accretion may become prohibitive, especially if the 
lifetimes of {\it circumbinary} disks are shorter than around 
single stars. The latter seems to be the case at least for
the {\it circumstellar} disks in binaries (Cieza \etal 2009; 
Kraus \etal 2012). 

Different ideas were proposed to alleviate this planetesimal 
fragmentation issue. Orbital alignment due to gas drag 
in presence of secular forcing has been suggested to
facilitate growth of $1-10$ km bodies (Scholl \etal 2007),
but this effect is size-dependent and does not work well
for a broad spectrum of planetesimal masses. Paardekooper 
\etal (2012) explored very efficient accretion of dust by
growing planetesimals as another way of bypassing the 
fragmentation problem. 

In this work we show that gravitational effect of the
circumbinary disk, in which planetesimals are immersed, 
modifies their secular excitation and significantly 
lowers their eccentricities beyond 2-4 AU from the 
star. Similar mechanism was invoked in Rafikov (2012b)
to explain the origin of circumstellar planets on wide orbits 
($\sim 2$ AU) in small separation binaries, $a_b\approx 20$ 
AU. A unique feature of the circumbinary planet formation
is that the effect of the disk gravity on the central binary 
often turns out being more important than the direct effect 
of the disk on planetesimal orbits. We now describe 
this idea in more detail.


\section{Secular evolution.}  
\label{sect:dynamics}


We consider motion of massless planetesimals around a 
central binary. Binary has components 
with masses $M_p$ and $M_s<M_p$ (mass ratio 
$\mu\equiv M_s/M_b<1$, where $M_b=M_p+M_s$ is
the binary mass), its semimajor axis and eccentricity 
are $a_b$ and $e_b$ correspondingly. For simplicity we will 
assume the disk to be axisymmetric with respect to the binary 
barycenter, i.e. the disk surface density is $\Sigma(r)$,
where ${\bf r}$ is the distance from the barycenter.
Since we are primarily interested in the effect of the binary 
on planetesimal dynamics we neglect gas drag in this work 
(see \S \ref{sect:planet_formation}). Planetesimals start 
on circular orbits and we are interested in their 
eccentricity evolution driven by the time-dependent 
and non-axisymmetric potential of the binary.

Following MN04 we write down the equation 
of planetesimal motion as
\ba
\frac{d^2{\bf r}}{dt^2}= -\frac{G(M_p+M_s)}{r^3}{\bf r}
+\nabla R,
\label{eq:eom}
\ea
where the disturbing function $R$ is 
\ba 
R=\frac{GM_p}{|{\bf r}-{\bf r}_p|}+\frac{GM_s}{|{\bf r}-{\bf r}_s|}
-\frac{G(M_p+M_s)}{r}-U_d.
\label{eq:dist_f}
\ea
Here $U_d$ is the disk potential and ${\bf r}_p$ and ${\bf r}_s$
are the vectors to primary and secondary from the barycenter of the 
binary.
 
Using Murray \& Dermott (1999) we expand the disturbing 
function to second order in planetesimal 
and binary eccentricities $e$ and $e_b$, retaining terms up to 
$O(e^2)$ and $O(e e_b)$, and then average it over the 
binary and planetesimal mean longitudes, thus eliminating 
short-period perturbations. We additionally expanded Laplace 
coefficients assuming $a_b/a\ll 1$, where $a$ 
is the planetesimal semi-major axis. The resulting 
secular disturbing function $R^{sec}$ is 
\ba
R^{sec}&=&\frac{1}{2}a^2 n\dot 
\varpi_d e^2+\frac{\mu(1-\mu)}{4}n_b^2a^2\left(\frac{a_b}{a}\right)^5
\nonumber\\
&\times &\left[\frac{3}{2}e^2+\frac{15}{4}(1-2\mu)\frac{a_b}{a} 
e e_b \cos(\varpi-\varpi_b)\right],
\label{eq:sec}
\ea
where we dropped insignificant $e$-independent terms. 
Here $n_b=(GM_b/a_b^3)^{1/2}$ is the binary mean motion,
\ba
\dot\varpi_d=-\frac{1}{2nr^2}
\frac{\partial}{\partial r}\left(r^2
\frac{\partial U_d}{\partial r}\right)\Big|_{r=a}
\label{eq:disk_prec}
\ea
is the precession frequency of planetesimal orbit due to
the disk potential, and $n=(GM_b/a^3)^{1/2}$ 
is the mean motion around a point mass $M_p+M_s$. To 
lowest order in $e$ these expressions coincide with 
secular expansion of MN04, if we set $\dot\varpi_d=0$.

\begin{center}
\begin{deluxetable}{lrrrrrr}
\tablecolumns{8}
\tablecaption{Circumbinary planetary systems
\label{table2}}
\tablehead{
\colhead{System}&
\colhead{$a_b$}&
\colhead{$e_b$}&
\colhead{$M_p$}&
\colhead{$M_s$}&
\colhead{$a_{pl}$}&
\colhead{$a_{\rm form}\tablenotemark{a}$}\\
\colhead{}&
\colhead{(AU)}&
\colhead{}&
\colhead{(M$_\odot$)}&
\colhead{(M$_\odot$)}&
\colhead{(AU)}&
\colhead{(AU)}
}
\startdata
Kepler-16\tablenotemark{1} & 0.22 & 0.16 & 0.69 & 0.2 & 0.7 & 2.3-4.4 \\
Kepler-34\tablenotemark{2} & 0.23 & 0.52 & 1.05 & 1.02 & 1.1 & 2.1-4.3 \\
Kepler-35\tablenotemark{2} & 0.18 & 0.14 & 0.89 & 0.81 & 0.6 & 1.7-3.5 \\
Kepler-38\tablenotemark{3} & 0.15 & 0.1 & 0.95 & 0.25 & 0.46 & 1.9-3.5 \\
Kepler-47\tablenotemark{4} & 0.084 & 0.02 & 1.04 & 0.36 & 0.29\tablenotemark{b} & 1.0-1.9 \\
KIC 4862625\tablenotemark{5} & 0.17 & 0.21 & 1.38 & 0.39 & 0.63 & 2.7-5.1 \\
\enddata
\tablenotetext{}{\tablenotemark{1}Doyle \etal (2011); \tablenotemark{2}Welsh \etal (2012); \tablenotemark{3}Orosz \etal (2012b); \tablenotemark{4}Orosz \etal (2012a); \tablenotemark{5}Schwamb \etal (2012)}
\tablenotetext{a}{Inner edge of accretion-friendly zone, see 
\S \ref{sect:planet_formation}}
\tablenotetext{b}{Semi-major axis of the inner planet}
\end{deluxetable}
\end{center}


\subsection{Evolution equations.}  
\label{sect:evolution}

We now follow standard procedure (Murray \& Dermott 1999) 
and introduce eccentricity vector 
${\bf e}=(k,h)=(e\cos\varpi,e\sin\varpi)$.
Defining
\ba
A &=& \frac{3}{4}\mu(1-\mu)\frac{n_b^2}{n}
\left(\frac{a_b}{a}\right)^5,
\label{eq:defs1}\\
B &=& \frac{15}{16}\mu(1-\mu)(1-2\mu)\frac{n_b^2}{n}
\left(\frac{a_b}{a}\right)^6 e_b,
\label{eq:defs2}
\ea
we can write
\ba
\frac{R}{na^2}=\frac{A+\dot\varpi_d}{2}
\left(h^2+k^2\right)
+B(k\cos\varpi_b+h\sin\varpi_b).
\label{eq:R_reduced}
\ea
This expression is accurate to $O(e_b^2)$. Evolution equations 
for $h$ and $k$ (Murray \& Dermott 1999) attain a relatively 
simple form 
\ba
\frac{dh}{dt}&=& (A+\dot\varpi_d)k+B\cos\varpi_b,
\label{eq:evol_eqs_h}\\
\frac{dk}{dt}&=&-(A+\dot\varpi_d)h-B\sin\varpi_b.
\label{eq:evol_eqs_k}
\ea
In the disk-free case ($\dot\varpi_d=0$) one recovers the
secular evolution equations from MN04.

Now we introduce an important modification to the setup used 
in MN04. We assume that $\varpi_b$ is not constant
but {\it linearly increases with time} at constant rate
$\dot\varpi_d$, which we specify later in \S 
\ref{sect:disk}, i.e. $\varpi_b=\dot\varpi_d t$. 
This makes forcing term in equations (\ref{eq:evol_eqs_h}) and 
(\ref{eq:evol_eqs_k}) time-dependent, but still permits 
analytical solution in the form  
${\bf e}(t)={\bf e}_{\rm free}(t)+{\bf e}_{\rm forced}(t)$,
where 
\ba
\left\{
\begin{array}{l}
k_{\rm free}(t)\\
h_{\rm free}(t)
\end{array}
\right\}
=e_{\rm free}
\left\{
\begin{array}{l}
\cos\left[(A+\dot\varpi_d)t+\varpi_0\right]\\
\sin\left[(A+\dot\varpi_d)t+\varpi_0\right]
\end{array}
\right\}
\label{eq:free_part}
\ea
and 
\ba
\left\{
\begin{array}{l}
k_{\rm forced}(t)\\
h_{\rm forced}(t)\\
\end{array}
\right\}
&=& -e_{\rm forced}
\left\{
\begin{array}{l}
\cos\varpi_b\\
\sin\varpi_b
\end{array}
\right\},
\label{eq:forced_part}
\\
e_{\rm forced} &=& \frac{B}{A+\dot\varpi_d-\dot\varpi_b}
\label{eq:forced_part_ampl}
\ea
Thus, free eccentricity vector ${\bf e}_{\rm free}$ rotates at 
a rate $A+\dot\varpi_d$ around the endpoint of vector 
${\bf e}_{\rm forced}$, which itself {\it rotates 
about the origin} with the rate $\dot\varpi_b$.
Setting $\dot\varpi_d=\dot\varpi_b=0$ brings us back to the 
MN04 solution.

Planetesimals starting on circular orbits have 
$e_{\rm free}=e_{\rm forced}$ and reach the maximum 
eccentricity of 
\ba
e=2e_{\rm forced}=\frac{2B}{A+\dot\varpi_d-\dot\varpi_b} 
\label{eq:e}
\ea
in the course of their secular evolution.


\section{Disk model.}  
\label{sect:disk}

Structure of circumbinary disks is different from that
of protoplanetary disks around single stars. As a result of 
viscous evolution the latter are expected to have mass
accretion rate $\dot M$ independent of radius. In the 
case of circumbinary disk the torque due to binary stops the
inward flow of matter and truncates the disk at inner 
radius $r_{in}$, interior to which $\Sigma$ and $\dot M$ are small.
Simulations find that $r_{in}\approx 2a_b$ for binaries 
with mass ratio $M_s/M_p\sim 1$ (MacFadyen \& Milosavljevic 2008).

Injection of angular momentum at its inner edge causes the
disk to evolve into a configuration, in which the viscous
angular momentum flux 
\ba
F_J=3\pi\nu\Sigma\Omega r^2,
\label{eq:F_J}
\ea
rather than $\dot M$, is constant with radius (Pringle 1991; 
Ivanov \etal 1999; Rafikov 2012a). Detailed description of 
circumbinary protoplanetary disk properties will be provided 
elsewhere (Garmilla \& Rafikov, in preparation); for the 
purposes of this paper we will assume that the disk is passive, 
i.e. heated predominantly by the combined light of the binary
components. Assuming $\alpha$-model for the viscosity $\nu$
one finds
\ba
\Sigma=\frac{F_J}{3\pi\alpha c_s^2 r^2},
\label{eq:sig}
\ea 
where $c_s$ is a sound speed. If $\alpha$ is independent of 
radius, and midplane disk temperature scales as 
$T(r)\propto r^{-k}$, then the constant $F_J$ disk has 
density profile 
\ba
\Sigma(r)=\Sigma_{in}\left(\frac{r_{in}}{r}\right)^p,~~~p=2-k,
\label{eq:sig1}
\ea
where $\Sigma_{in}\equiv\Sigma(r_{in})$. 
Passive disks typically have $k\approx 1/2$, in particular 
Chiang \& Goldreich (1997) find $k=3/7$. For this reason we
will take $p=3/2$ in this work, which is similar to the $\Sigma$
slope of the Minimum Mass Solar Nebula (Hayashi 1981), and is 
different from $p\approx 1$ expected for a passive constant 
$\dot M$ disk (Rafikov 2012b).

Most of the mass in a constant $F_J$ disk is contained in
its outer regions. Assuming that $p=3/2$ profile is maintained
all the way to the outer radius $r_o$ we find that 
\ba
\Sigma(r)&\approx &\frac{M_d}{4\pi r_o^{2}}
\left(\frac{r_o}{r}\right)^{3/2}
\label{eq:sig2}\\
&\approx & 1.3\times 10^3~\mbox{g cm}^{-2}
\frac{M_d}{0.01M_\odot}r_{o,30}^{-1/2}r_1^{-3/2},
\ea
where $M_d$ is the disk mass, $r_{o,30}=r_o/(30$AU), 
$r_{1}=r/(1$AU).

We show in Appendix \ref{app} that disk with density profile 
(\ref{eq:sig2}) gives rise to apsidal precession of
planetesimal orbits at the rate
\ba
\dot\varpi_d=-\frac{\pi K_d}{4} \frac{G\Sigma(r)}{rn}
\approx -\frac{K_d}{16}n\frac{M_d}{M_b}
\left(\frac{r}{r_o}\right)^{1/2},
\label{eq:omega_p}
\ea
(where $K_d\approx 4.4$), as well as apsidal precession of 
the central binary at the rate 
\ba
\dot\varpi_b &=& \frac{\pi G\Sigma_{in}}{n_b r_{in}}\tilde\phi
\left(\frac{a_b}{r_{in}}\right)=\frac{\tilde\phi}{4}n_b
\frac{M_d}{M_b}\frac{a_b^3}{r_o^{1/2}r_{in}^{5/2}},
\label{eq:omega_b}
\ea
where $\tilde \phi\approx 0.5$, see equation (\ref{eq:phi_series}).

Equation (\ref{eq:omega_b}) assumes disk to be sharply truncated 
at $r_{in}$, while in reality $\Sigma$ smoothly (but rapidly) 
goes to zero near $r_{in}$ (MacFadyen \& Milosavljevic 2008), 
which lowers the amount of mass near the binary and may reduce 
$\dot\varpi_b$. To account for this we will sometimes 
consider wider disk cavity, e.g. $r_{in}=3a_b$ (Pelupessy \& 
Portegies Zwart 2012) instead of the more conventional $2a_b$.

\begin{figure}
\plotone{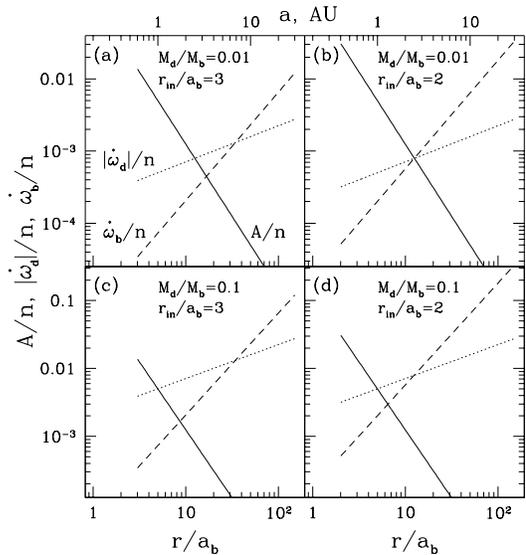}
\caption{
Characteristic precession frequencies of the problem as a function
of distance from the binary (in units of $a_b$ on the lower axis 
and in AU on the upper), for different values of the relative 
disk mass $M_d/M_b$ and the size of the inner disk cavity 
$r_{in}/a_b$, as labeled on panels. Calculation assumes 
$M_b=M_\odot$, $\mu=0.21$, $a_b=0.2$ AU, and $e_b=0.2$.
Shown are $A/n$ ({\it solid}), 
$|\dot\varpi_d|/n$ ({\it dotted}), and $\dot\varpi_b/n$ 
({\it dashed}), marked in panel (a). Note that for small
$M_d/M_b$ and $r_{in}/a_b$ region where $|\dot\varpi_d|$ 
dominates over other frequencies disappears.  
\label{fig:freqs}}
\end{figure}


\section{Planetesimal velocities.}  
\label{sect:excitation}

Equation (\ref{eq:forced_part_ampl}) shows that planetesimal 
eccentricity is determined by $A$, $\dot\varpi_d$ and 
$\dot\varpi_b$. We plot the behavior of these frequencies as 
a function of $r$ in Figure \ref{fig:freqs}. Using equations 
(\ref{eq:defs1}), (\ref{eq:omega_p}) and (\ref{eq:omega_b}) 
we find that $\dot\varpi_b/A\propto (r/a_b)^{7/2}$ and
$\dot\varpi_b/|\dot\varpi_d|\propto r/a_b$. Both ratios increase 
with $r$ so that $\dot\varpi_b\gtrsim A$ for $r\gtrsim r_A$, 
where
\ba
\frac{r_A}{a_b}&\approx & 14\left[\frac{\mu(1-\mu)}{0.25}
\frac{0.01}{M_d/M_b}\right]^{2/7}
\nonumber\\
&\times &\left(\frac{r_{o,30}}{a_{b,0.2}}\right)^{1/7}
\left(\frac{r_{in}/a_b}{2}\right)^{5/7},
\label{eq:r_A}
\ea
with $a_{b,0.2}\equiv a_b/(0.2$AU). Also, 
$\dot\varpi_b\gtrsim |\dot\varpi_d|$ for 
$r\gtrsim r_d$, where (setting $K_d=4.4$ and
$\tilde\phi=0.5$)
\ba
\frac{r_d}{a_b}&\approx & 12\left(\frac{r_{in}/a_b}{2}\right)^{5/2}.
\label{eq:r_d}
\ea

Thus, for a particular 
set of parameters adopted in these estimates one finds that 
$\dot\varpi_b$ dominates the behavior of planetesimal 
eccentricity beyond $(10-20)a_b$. Depending on $M_d/M_b$
and $r_{in}/a_b$ there may exist intermediate region 
around $r\sim 10a_b$, where relative precession of the 
planetesimal and binary orbits is dominated by 
$\dot\varpi_d$. Such region disappears for lower $M_d$
and smaller inner cavity size $r_{in}$, see Figure  
\ref{fig:freqs}b.

\begin{figure}
\plotone{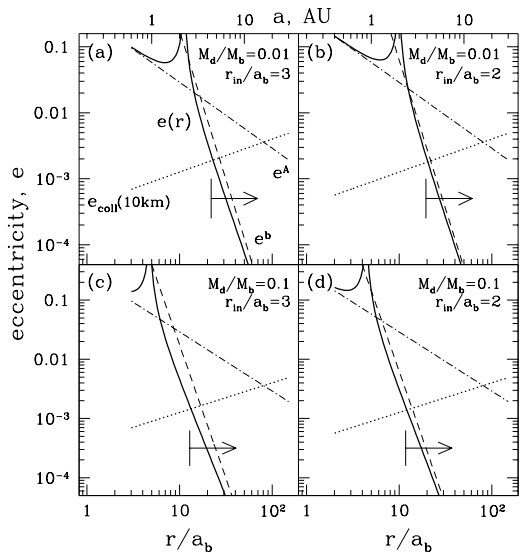}
\caption{
Maximum planetesimal eccentricity $e(r)$ (solid curve,
Eq. [\ref{eq:e}]) as a function of radius, for different
values of $M_d/M_b$ and $r_{in}/a_b$. Dot-dashed and 
dashed lines illustrate asymptotic behavior of 
eccentricity given by $e^A$ (Eq. [\ref{eq:eA}]) and
$e^b$ (Eq. [\ref{eq:eb}]). All these curves scale linearly 
with the binary eccentricity $e_b$, assumed equal to $0.2$
in this calculation (also $M_b=M_\odot$, $\mu=0.21$, 
$a_b=0.2$ AU). 
Dotted line is the eccentricity 
$e_{coll}$ (Eq. [\ref{eq:e_coll}]) below which 10 km objects
(density $\rho=3$ g cm$^{-3}$) can grow in collisions, 
according to the criterion 
(\ref{eq:cond}). Planetesimal growth is unimpeded by
fragmentation when $e(r)<e_{coll}(r)$ (region 
to the right from vertical bars with arrows).  
\label{fig:eccs}}
\end{figure}

In Figure \ref{fig:eccs} we illustrate radial dependence 
of the characteristic planetesimal eccentricity $e(r)$ computed with 
equation (\ref{eq:e}). One feature that is obvious in these 
plots is the secular resonance at $\sim 1-2$ AU,
where $A$ is equal to $\dot\varpi_b+|\dot\varpi_d|$ and
our solution presented in \S \ref{sect:evolution} breaks 
down. This region of enhanced excitation is quite narrow,
and just outside of it $e(r)$ rapidly declines with $r$.

We also show in this Figure asymptotic behavior of eccentricity 
$e^A(r)$ found when the relative precession of planetesimal 
and binary orbits is dominated by the {\it potential of the 
secondary} ($A\gtrsim |\dot\varpi_d|,\dot\varpi_b$), see 
(\ref{eq:defs1}), (\ref{eq:defs2}):
$e\to e^A=2B/A$, where (MN04)
\ba
e^A=\frac{5(1-2\mu)}{2}\frac{a_b}{r}e_b\approx 
0.02(1-2\mu)\frac{25}{r/a_b}\frac{e_b}{0.2},
\label{eq:eA}
\ea
and $r/a_b=25$ is chosen 
so that $r=5$ AU if $a_b=0.2$ AU. 

When relative precession is dominated by the {\it binary 
precession} ($\dot\varpi_b\gtrsim A,|\dot\varpi_d|$), 
equation (\ref{eq:omega_b}) predicts 
$e\to e^b=2B/\dot\varpi_b$, where 
\ba
e^b &=&\frac{15\psi(\mu)}{2\tilde\phi}\frac{M_b}{M_d}
\left(\frac{r_o}{a_b}\right)^{1/2}
\left(\frac{r_{in}}{a_b}\right)^{5/2}
\left(\frac{a_b}{r}\right)^{9/2}e_b
\label{eq:eb}\\
&\approx & 10^{-3}\frac{\psi(\mu)}{0.1}\frac{0.01}{M_d/M_b}
\left(\frac{r_{o,30}}{a_{b,0.2}}\right)^{1/2}
\nonumber\\
&\times &\left(\frac{r_{in}/a_b}{2}\right)^{5/2}
\left(\frac{r/a_b}{25}\right)^{-9/2}\frac{e_b}{0.2}.
\ea
Here $\psi(\mu)\equiv\mu(1-\mu)(1-2\mu)$ and
for $0<\mu<0.5$ the maximum value of $\psi\approx 0.096$ 
is achieved at $\mu\approx 0.21$.

Between these two limits, at $r\sim 1$ AU, disk-driven 
planetesimal precession may dominate (see Figure 
\ref{fig:freqs}), but usually marginally. For this reason 
we do not show asymptotic scaling for this regime.

At large separations, of order several AU, $e(r)\to e^b(r)$ 
and falls off very steeply with $r$. At these 
separations the simple formula (\ref{eq:eA}) not accounting 
for the gravitational effects of the disk overestimates 
planetesimal eccentricity by more than an order of magnitude.


\section{Short-period eccentricity variations.}  
\label{sect:short}

Planetesimal motion is affected not only by the explicitly 
time-independent, secular part of the disturbing function 
(\ref{eq:sec}), but also by the short-term perturbations 
varying on timescales $\sim n_b^{-1}$ and $n^{-1}$. The 
former average out to zero over the
planetesimal orbital motion, but the latter were suggested 
to affect planetesimal dynamics. 

In particular, MN04 and Paardekooper 
\etal (2012) found that even for circular 
binaries ($e_b=0$), when secular excitation is absent (see 
equations [\ref{eq:defs2}] and [\ref{eq:e}]), time-dependent  
contributions to the disturbing function varying on the 
planetesimal orbital timescale $n^{-1}$ (i.e. averaged over 
the {\it binary} period) still result in eccentricity evolution.
This result is surprising since in $e_b=0$ case the potential 
of the binary averaged over the fast binary orbital timescale 
is time-independent and axisymmetric. Consequently, both energy 
and angular momentum of the planetesimal must be conserved 
precluding its eccentricity evolution.  

The unexpected finding of MN04 and 
Paardekooper \etal (2012) is most likely related to their 
choice of a reference circular orbit and osculating orbital 
elements. The former was defined as the circular 
orbit in a Keplerian potential for the mass $M_b=M_p+M_s$. 
However,
one can show that the true axisymmetric part of the binary
(plus disk)  potential to lowest order in $a_b/r$ is
\ba
U_{m=0}(r) &=& U_d(r)-\frac{GM_b}{r}
\nonumber\\
& \times &\left[1+
\frac{1}{4}\mu(1-\mu)\left(\frac{a_b}{r}\right)^2
\left(1+\frac{3}{2}e_b^2\right)\right].
\label{eq:axisym}
\ea
Circular orbits in this potential have {\it higher} circular
speed than orbits in a Keplerian potential for mass $M_b$ at 
the same distance. Thus, a particle initialized on ``circular''
orbit, assuming $-GM_b/r$ potential, is in fact started at
the {\it apoapse of eccentric orbit} in the true potential 
(\ref{eq:axisym}). Not surprisingly, its orbit will remain
eccentric, with eccentricity determined ultimately by the 
difference between the circular velocities in $U_{m=0}$ and 
the $M_b$ point mass potentials. It is trivial to show
that this fictitious eccentricity is 
$(3/4)\mu(1-\mu)(a_b/r)^2$, which is in perfect agreement 
with the numerical calculations of MN04 and analytical 
result Paardekooper \etal (2012) for 
$e_b=0$. The importance of proper definition of osculating
orbital elements has been previously emphasized by 
Marzari \etal (2008).

Careful analysis in the $e_b\neq 0$ case shows that the 
disturbing function contains terms varying on planetesimal 
orbital timescale, which formally result in eccentricity of 
order $\mu(1-\mu)(a_b/r)^2e_b^2$. Even taken at face value, 
this eccentricity is most likely too small to affect the results 
in \S \ref{sect:excitation}; whether it produces significant 
{\it relative} velocity between colliding objects is even 
less obvious. We leave the detailed study of the short-term
eccentricity variations for the future.


\section{Implications for planet formation.}  
\label{sect:planet_formation}

We now address the issue of fragmentation barrier for 
planetesimal growth. Based on work of Leinhardt \& Stewart 
(2012) we estimate that a gravity-dominated body (e.g. a 
rubble pile) of radius $d$ survives in a collision with 
a body of {\it equal size} whenever the collision velocity 
$v_{coll}$ (at large separation) satisfies a simple 
condition (Rafikov 2012b)
\ba
v_{coll}\lesssim 2v_{esc},
\label{eq:cond}
\ea
where the 
escape speed from the surface of an object of radius $d$ 
and bulk density $\rho$ is $v_{esc}=[(8\pi/3)G\rho]^{1/2}d$. 
This condition is similar to the one used in MN04. It can 
be translated into the constraint on the maximum planetesimal
eccentricity $e_{coll}\approx 2v_{esc}/v_K$ (where $v_K$ is the 
Keplerian speed), at which an equal-mass collision does not
result in the net loss of mass:
\ba
e_{coll}(r,d)&\approx &\left(\frac{32\pi}{3}\frac{\rho r d^2}
{M_b}\right)^{1/2}
\nonumber\\
&\approx & 2\times 10^{-3}
\left(\frac{\rho_3}{M_{b,1}}
\frac{r}{5\mbox{AU}}\right)^{1/2}d_{10},
\label{eq:e_coll}
\ea
where $M_{b,1}\equiv M_b/M_\odot$, 
$\rho_3\equiv\rho/(3$ g cm$^{-3})$,  
and $d_{10}\equiv d/(10$km).

We will now assume that if (1) the characteristic planetesimal 
eccentricity given by equation (\ref{eq:e}) is below 
$e_{coll}(r,d_s)$ at some distance $r$ and (2) planetesimals are
strength-dominated below the radius $d_s$, then fragmentation 
barrier at this separation $r$ is bypassed. 

To be specific, we take $d_s=10$ km in our study. This 
may seem somewhat large since collisionally assembled objects 
may be rubble piles. On the other hand, more
sophisticated fragmentation criteria accounting for the size 
spectrum of colliding objects (i.e. incorporating more than 
just equal-mass collisions) typically find our 
survival criterion (\ref{eq:cond}) too restrictive (Th\'ebault 
2011; Rafikov 2012b). This may justify relatively large value 
of $d_s=10$ km.

Figure \ref{fig:eccs} shows that with this value of $d_s$ 
fragmentation barrier does not get in the way of planetesimal 
growth at separations $\gtrsim 2-4$ AU. Inner radius of 
the accretion-friendly zone $a_{\rm form}$ depends mainly 
on the size of the inner cavity in the disk and to some 
extent on the disk mass. A rough estimate of $a_{\rm form}$
(typically an overestimate by up to a factor of 2) can be 
obtained by equating $e^b$ and $e_{coll}$:
\ba
a_{\rm form}&\approx & 22a_b\left(\frac{M_{b,1}r_{o,30}}{\rho_3}
\right)^{1/10}
\left(\frac{r_{in}/a_b}{2}\right)^{1/2}
\nonumber\\
&\times &
\left(\frac{\psi(\mu)}{0.1}\frac{0.01}{M_d/M_b}
\frac{e_b}{0.2}d_{10}^{-1}a_{b,0.2}^{-1}\right)^{1/5}.
\label{eq:a_form}
\ea
Larger $r_{in}/a_b$ means less disk mass near the binary 
and slower binary precession. Larger $M_d$ accelerates 
disk-induced precession of both the planetesimal orbit 
and the binary. Larger value of critical planetesimal 
size $d_s$ also shifts $a_{\rm form}$ closer to the binary. 

In the absence of disk-driven precession, using $e^A$ instead 
of $e^b$ in the growth condition (\ref{eq:cond}) to calculate 
$a_{\rm form}$, we would obtain 
$a_{\rm form}\approx 90a_b\approx 17$ AU (crossing of dotted and 
dot-dashed lines in Figure \ref{fig:eccs}), pushing 
accretion-friendly zone much further from the binary, 
in agreement with MN04. Formation of cores massive enough to trigger 
core accretion ($\sim 10M_\oplus$) prior to disk dispersal 
is more problematic at this separation than at 2-3 AU, because of the 
longer dynamical timescale. Thus, by extending inward the region, 
where planetesimals can grow effectively, the gravitational effect 
of the disk on planetesimal secular evolution facilitates 
circumbinary planet formation via the reduction of the planetary
accretion timescale.

In Table \ref{table2} we show $a_{\rm form}$ computed using
parameters of actual {\it Kepler} circumbinary planetary 
systems. Minimum and maximum values shown correspond to 
$M_d/M_b=0.1$, $r_{in}/a_b=2$ and $M_d/M_b=0.01$, 
$r_{in}/a_b=3$, correspondingly. 
In Kepler-47 system\footnote{Equation (\ref{eq:a_form}) 
works poorly for this system because of large 
$\dot\varpi_d$.} secular excitation is 
additionally suppressed because $e_b=0.02\ll 1$ 
(Orosz \etal 2012a). In Kepler-34 and Kepler-35 it is reduced 
because $\mu$ is very close to $0.5$: $\mu=0.493$
and $0.477$, correspondingly (Welsh \etal 2012). 
Compared to the nominal $\mu=0.2$ case, we
find $e^A$ in these systems to be suppressed by $\approx 40$
and $\approx 13$, respectively, while $e^d$ and $e^b$ are
lowered by $\approx 27$ and $9$, correspondingly. This  
shifts accretion-friendly zone closer to the binary. 

Nevertheless, even accounting for the gravitational 
effect of the disk, in-situ formation still does not seem 
viable for the {\it Kepler} circumbinary planets, which have 
$a_{\rm form}>a_{pl}$ (Table \ref{table2}). Thus, some form 
of inward migration is still needed to deliver these 
planets to their current locations.

We also note that gas drag is not important for resolving
the fragmentation barrier issue in Kepler circumbinary 
systems: fast relative precession of planetesimal and binary 
orbits (mainly due to $\dot\varpi_b\gg A$) makes planetesimal 
apsidal alignment inefficient,
contrary to standard expectation without disk gravity 
(Scholl \etal 2007). Calculations similar to the one in 
Rafikov (2012b) demonstrate that gas drag regulates 
eccentricity behavior at 2-3 AU only for bodies smaller 
than 1 km, which is below our adopted threshold size $d_s$.

However, gas drag may have detrimental effect on circumbinary 
planet formation if the turbulence (Meschiari 2012b) or 
the density waves driven by the binary in the disk 
(Marzari \etal 2008) can strongly excite planetesimal 
eccentricities at large $r/a_b$. Drag-induced inspiral of 
solids is another possible obstacle for planet 
formation. We leave detailed exploration of these issues 
to future study.


\acknowledgements

This work was supported by NSF via grant AST-0908269.



\appendix


\section{Disk-induced precession rates.}  
\label{app}

Potential $U_d$ of an axisymmetric disk with a power-law surface 
density profile in the form (\ref{eq:sig1}), truncated at the 
inner radius $r_{in}$ (i.e. $\Sigma(r)=0$ for $r<r_{in}$)
is given by the following expression far from the inner 
edge of the disk, $r\gg r_{in}$ (Ward 1981): 
\ba
U_d(r) \approx  -2\pi K_d G \Sigma_{in}
\left(\frac{r_{in}}{r}\right)^p r,~~~~~~K_d = 
\sum\limits_{l=0}^\infty\frac{A_l(4l+1)}
{(2l+p-1)(2l-p+2)}\approx 4.4,
\label{eq:disk_pot_far}
\ea
where $A_l=\left[(2l)!/(2^{2l}(l!)^2)\right]^2$, and the numerical 
value of $K_d$ is for $p=3/2$. Substituting this expression 
into equation (\ref{eq:disk_prec}), assuming $p=3/2$ and
using equation (\ref{eq:sig2}), we arrive at the result 
(\ref{eq:omega_p}).

To calculate apsidal precession of the binary we
use the expression for $U_d$ inside 
the inner edge ($r<r_{in}$) of a sharply truncated disk 
(Ward 1981): 
\ba
U_d(r)=-2\pi G\Sigma_{in}r_{in}
\phi\left(\frac{r}{r_{in}}\right),~~~~~\phi(z)=
\sum\limits_{l=0}^\infty\frac{A_lz^{2l}}{2l+p-1}.
\label{eq:disk_pot_inside}
\ea
Taking into account that both stars move in this potential 
one can show using equation (\ref{eq:disk_prec}) that 
binary precesses at the rate (\ref{eq:omega_b}) with 
\ba
\tilde\phi(z) &= &\sum\limits_{l=1}^\infty A_l
\frac{2l(2l+1)}{2l+p-1}
\left[\mu^{2l}+(1-\mu)^{2l}\right]
z^{2(l-1)},
\label{eq:phi_series}
\ea
where equation (\ref{eq:sig2}) was used. We numerically found that 
$\tilde\phi\approx 0.3-0.8$ for $a_b/r_{in}=0.3-0.5$ and
$\mu=0-0.5$. For simplicity, in this study
we simply set $\tilde\phi\approx 0.5$.

\end{document}